\documentclass[twocolumn,floatfix,pra,aps,showpacs,superscriptaddress]{revtex4}
\usepackage{epsfig,graphicx}
\usepackage{flafter}
\usepackage{amsmath}
\usepackage{color}
\usepackage{bm}
\usepackage{amssymb}
\usepackage{epstopdf}
\usepackage{amsmath}
\usepackage{amstext}
\usepackage{amsthm}
\usepackage{amsfonts}
\usepackage{latexsym}

\newcommand{\ket}[1]{\left\vert#1\right\rangle}

\renewcommand{\emph}[1]{{\it #1}}
\renewcommand{\vec}[1]{\boldsymbol{#1}}

\begin{document}

\title{The role of anisotropy in dipolar bosons in triple-well potentials}

\author{A. Gallem\'{\i}}
\affiliation{Departament d'Estructura i Constituents de la Mat\`{e}ria, Universitat de Barcelona, E-08028
Barcelona, Spain}
\author{M. Guilleumas}
\author{R. Mayol}
\affiliation{Departament d'Estructura i Constituents de la Mat\`{e}ria, Universitat de Barcelona, E-08028 Barcelona, Spain}
\affiliation{Institut de Nanoci\`encia i Nanotecnologia de la Universitat de Barcelona, IN$\,^2$UB, E-08028 Barcelona, Spain}
\author{A. Sanpera}
\affiliation{Instituci\'o Catalana de Recerca i Estudis Avan\c{c}ats, ICREA, E-08011 Barcelona, Spain} 
\affiliation{Departament de F\'{\i}sica, Universitat
Aut\`{o}noma de Barcelona, E-08193 Bellaterra, Spain}

\date{\today}

\begin{abstract}
Mesoscopic samples of polarized dipolar atoms confined in three spatially separated traps conform an extended Bose-Hubbard 
Hamiltonian in which different quantum phases appear depending on the competition between tunneling, on-site and long 
range inter-site dipole-dipole interactions. 
Here, by choosing an appropriate configuration of triple-wells, we analyze the role played by the anisotropic character 
inherent to the dipolar interaction in the phase diagram of the system. We further characterize the different phases as 
well as their boundaries by means of their entanglement properties.
\end{abstract}

\pacs{03.75.Hh, 03.75.Lm, 03.75.Gg, 67.85.-d}

\maketitle

\section{Introduction}
The experimental achievement of dipolar Bose-Einstein Condensation first with chromium \cite{Griesmaier2005} 
and more recently with dysprosium~\cite{Lu2011} and erbium atoms~\cite{Aikawa2012}, together with the recent 
progress in trapping and cooling dipolar molecules~\cite{Deiglmayr2008,DeMiranda2011}, 
has busted the physics of dipolar gases into various research directions. Dipolar Bose-Einstein condensates 
(dBECs) in the weakly interacting regime exhibit a strong dependence on the trap geometry~\cite{Lahaye2009,Abad2010}.
The dipolar effects in the strongly correlated regime are predicted to display dramatic differences both in static and dynamic properties.

Often, the strongly correlated regime is addressed by assuming dipolar atoms loaded in optical lattices. In such setups, the description of the dipolar gas requires an extended Bose-Hubbard model which predicts both insulating and superfluid exotic quantum phases~\cite{Lahaye2009}. 
The effect of dipolar interaction can also be explored in an intermediate regime consisting in just 
few traps hosting mesoscopic dipolar condensates, very much in the way Josephson physics has been studied in double-well potentials with cold gases \cite{Albiez2005,Levy2007}. The ground states of the latter present, in some region of the space parameters, 
quantum correlations that demand to go beyond a mean field description \cite{Zin2008,Javanainen1986,Buonsante2012,JuliaDiaz2010,Sakmann2009}. 
The physics of dipolar gases in double-well potentials reduces to the non-dipolar case 
since  the dipole-dipole interaction just renormalizes the contact one \cite{Abad2011}.

Therefore, in order to explore dipolar effects in this regime at least three sites are required. This kind of setup has 
been previously considered in the literature  \cite{Lahaye2010,Zhang2012,Peter2012,Dell'Anna2013}. For 
aligned triple-well potentials loaded with dBECs, the long-range character of the dipolar interaction yields several
mesoscopic quantum phases \cite{Lahaye2010}. 
Also, genuine multipartite entanglement 
appears in a triangular trap configuration with isotropic inter-site dipole-dipole interaction~\cite{Dell'Anna2013}. 
This case leads again to a renormalized 
contact interaction and effectively reduces to a non-dipolar situation.

Here, we consider dipolar gases in triangular traps with different dipole orientations.
For each configuration, we derive the phase diagram of the system and analyze their associated phase transitions. In this way we unveil the role played 
by the anisotropic character of the dipole-dipole interaction in the  ground state of the system. 
We further 
characterize the different phases as well as their boundaries by means of their entanglement properties, showing their relevance to 
understand subtle effects that discriminate between phase transitions and crossovers. 

The paper is organized as follows: in Sec. II, we introduce the system under study and review the derivation of the three-site dipolar 
Bose-Hubbard Hamiltonian. Sec. III is devoted to the analysis of the phase diagram, first in the atomic limit ({\it i.e.} for zero tunneling) and then 
with exact diagonalization of the Hamiltonian. In Sec. IV, we further characterize the different ground state phases encountered by their 
entanglement properties. Aside from the von Neumann entropy we also calculate the entanglement spectrum which permits to understand in a 
more quantitative way ground state properties. Our conclusions are presented in Sec. V.

\section{Three-site Bose-Hubbard Hamiltonian}

\begin{figure}[h!]
\centering
\includegraphics[width=\linewidth]{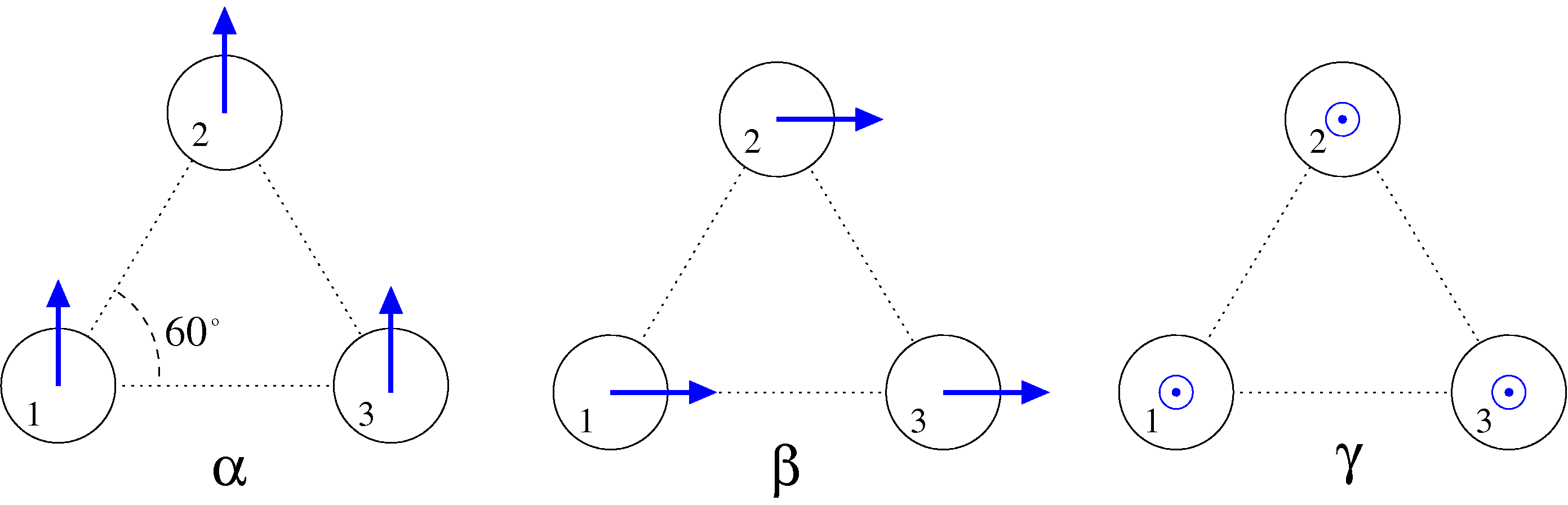}
\caption{
Polarization directions in the triangular lattice: $\alpha$-configuration with all dipoles along the vertical axis ($U_1 >0$), $\beta$-configuration with all dipoles along the horizontal axis ($U_1<0$) and $\gamma$-configuration with dipoles along the normal axis ($U_1>0$).
}
\label{Fig1}
\end{figure}

We assume $N$ dipolar bosons all polarized along the same direction. They are confined in a triple-well potential of depth $V_{0}$ arranged in an equilateral triangular configuration as schematically shown in Fig.$\,$\ref{Fig1}.  If $V_{0}$ is large compared to any other system's energy and the sites are sufficiently separated, the system can be described by three 
non-overlapping  wave functions, each localized on a well, $\phi_{i}(\vec{r})=\phi_{i}(\vec{r}-\vec{r_{i}})$,  for $i=1,2,3$. On each trap, $\phi_{i}(\vec{r})$ can be regarded as independent of the number of atoms $n_{i}$, if the latter is sufficiently small \cite{Lahaye2010,limitation}. The total number of atoms is conserved {\it i.e.} $N=\sum_{i}n_{i}$.

We define bosonic field operators that annihilate (create) a boson at a point $\vec{r}$ as 
$\hat{\psi}(\vec{r})=\sum_{i} \phi_{i}(\vec{r}) \, \hat{a}_{i}$, where, as usual,
$\hat{a}_{i} (\,\hat{a}_{i}^{\dagger})$ are the bosonic annihilation (creation) operators on site $i$ fulfilling canonical commutation relations. 
Under these assumptions the dipolar Bose-Hubbard (dBH) Hamiltonian reads:
\begin{align}
\mathcal{\hat H}=&-J\left[\hat{a}_1^\dagger \hat{a}_2 + \hat{a}_1^\dagger \hat{a}_3+ \hat{a}_2^\dagger \hat{a}_3 + h.c. \right]
+\frac{U_0}{2} \sum_{i=1}^{3} \hat{n}_i (\hat{n}_i-1) \nonumber \\
&+U_{1 2} \, \hat{n}_1 \hat{n}_2 + U_{2 3} \, \hat{n}_2 \hat{n}_3 + U_{1 3} \, \hat{n}_1 \hat{n}_3\,,
\label{hamiltonian-0}
\end{align}
where $\hat{n}_{i}=\hat{a}_{i}^{\dagger}\hat{a}_{i}$ is the particle number operator on  the 
$i$-th well.
The Hamiltonian (\ref{hamiltonian-0}) is characterized by three parameters: 
the tunneling rate ($J$), the on-site energy ($U_0$) and the inter-site energy ($U_{i j}$). 
The tunneling rate is given by:
\begin{equation}
J=-\int d^3 \vec r \, \phi_ i^*(\vec r) \left[ -\frac{\hbar^2}{2m}\vec \nabla^2 +V_{trap} (\vec r) \right] \phi_j(\vec r) \,,
\label{tunneling}
\end{equation}
with $i\neq j$. The on-site interaction includes both short-range and dipole-dipole contributions:
\begin{align}
U_0 & =  g  \int d^3 r \vert \phi_ i (\vec r) \vert ^{4} + \nonumber \\
&+\int d^3 \vec r \,\,d^3 \vec r\,' \,\vert \phi_ i (\vec r)\vert ^ 2 V_D(|\vec r-\vec r\,'|,\theta) \,\vert \phi_ i (\vec r\,')\vert^2 \,,
\label{onsite}
\end{align} 
where $V_D(|\vec r-\vec r\,'|,\theta)=d^2 \,(1-3\cos^2\theta)/|\vec r-\vec r\,'|^3\,$ is the dipolar interaction. Here $d^2=\mu_0 \mu_m^2/4\pi 
\,(d^2=\mu_e^2/4\pi\varepsilon_0)$, being $\vec {\mu_m} \,(\vec {\mu_e})$ the magnetic (electric) dipole moment, and $\theta$ the angle between the polarization direction and the relative position between two dipoles. 
The inter-site interaction  is a pure dipolar term given by:
\begin{eqnarray}
&U_{i j}& = \int d^3 \vec r \,d^3 \vec r'  \, |\phi_i (\vec r)|^ 2 \frac{d^{2}(1-3\cos^2\theta)}{|\vec r-\vec r\,'|^3} \,
          \, |\phi_j (\vec r')|^2 \nonumber\\
&=& (1-3\cos^{2}\theta_{i j}) \,U\,,
\label{intersite}
\end{eqnarray}
where we have considered an effective dipole for each site.
Assuming the dipoles $\vec{d}$ oriented along the coordinate axes we denote the possible configurations by
$\alpha$, $\beta$ and $\gamma$, as scketched in Fig.$\,$\ref{Fig1}.

Clearly, the orientation of the dipoles determines the inter-site interaction. Their values  are summarized in Table I for the different configurations. Notice that sites $1$ and $3$ are symmetric with respect to site $2$ implying that in all cases $U_{12}=U_{23}$.

\begin{table}[ht]
\centering
\begin{tabular}{| c | c | c | c | }
  \hline
 $U_{ij}/U$ & $\alpha$ & $\beta$ & $\gamma$ \\
\hline
\hspace{0.5cm} $U_{13}/U$ \hspace{0.5cm} & \hspace{0.5cm} $1$ \hspace{0.5cm} & \hspace{0.5cm} $-2$ \hspace{0.5cm} & \hspace{0.5cm} $1$ \hspace{0.5cm} \\ 
$U_{23}/U$ & $-5/4$ & $1/4$ & $1$ \\
 \hline 
 \end{tabular}
 \label{table1}
 \caption{Values of the inter-site dipole-dipole interaction for each configuration resulting from the anisotropy of the interaction.}
 \end{table}

In the $\alpha$-configuration, there is a repulsive interaction between sites $1$-$3$ but attractive otherwise. The $\beta$-configuration is the other way around, with attractive interaction between sites $1$-$3$ and repulsive otherwise. Finally, in the $\gamma$-configuration, the interaction becomes isotropic, repulsive and symmetric between all the modes.

Since the Hamiltonian commutes with the total number of particles $N$, Eq.(\ref{hamiltonian-0}) can be rewritten as: 
\begin{align}
\mathcal{\hat H}=&-J\left[\hat{a}_1^\dagger \hat{a}_2 + \hat{a}_1^\dagger \hat{a}_3+ \hat{a}_2^\dagger \hat{a}_3 + h.c. \right]\nonumber\\
&- U_0\left[ \hat{n}_2( \hat{n}_1 + \hat{n}_3)+ \hat{n}_1 \hat{n}_3 \right]\nonumber\\
&+ U_1 \left[ b\,(1-3\cos^2 \theta)\, \hat{n}_2( \hat{n}_1 + \hat{n}_3) + \hat{n}_1 \hat{n}_3 \right]\,,
\label{hamiltonian1}
\end{align}
where $U_{1}=U_{13}$ for each configuration and $\theta=\theta_{12}=\theta_{23}$. The parameter $b=U/U_{13}$ takes into account the factorization of $U_{13}$ and is equal to one for the $\alpha$ and $\gamma$-configurations, while $b=-1/2$ for the $\beta$-configuration as indicated in Table I. Moreover, we have neglected the constant term $U_0 N(N-1)/2$ in Eq.$\,$(\ref{hamiltonian1}) since it is a global energy shift.

\section{Ground states of the system: phase diagram}

The ground states of the dBH Hamiltonian depend on the parameters $J$, $U_0$ and $U_1$, as well
as on $N$, and different ground states belonging to different phases are expected to appear.  Despite the simplicity of the model, the competition  between tunneling, on-site and inter-site interactions in (\ref{hamiltonian1}) leads to some non trivial 
ground states. For $J\neq 0$, but small compared to $U_{0}$ and $U_{1}$, we expect some insulating phases, {\it i.e.\/} with a well defined number of bosons per site. To preserve symmetry conditions in the insulating phases, we fix the number of atoms to be even and multiple of 3. 

The structure of the dBH Hamiltonian makes it convenient to work in the Fock basis that labels the number of atoms in each well:
\begin{equation}
\vert \Psi \rangle = \sum_{n_1,n_2,n_3=0}^N \, C_{n_1,n_2,n_3} \vert n_1,n_2,n_3\rangle\,,
\label{psifock}
\end{equation}
where $C_{n_1,n_2,n_3}$ is the corresponding amplitude of the Fock state $\vert n_1,n_2,n_3\rangle$. 

The phase diagram has been found by exact diagonalization of Eq.$\,$(\ref{hamiltonian1}) 
setting the tunneling parameter $J/h=0.1 \mbox{Hz}$ and varying the interactions $-10\leq U_{0}/J\leq 10$ and $-10 \leq U_{1}/J \leq 10$.  In our simulations we fix the number of atoms to $N=12,\, 24\,,36$ and $48$. 
Results for larger number of atoms become computationally too time-consuming and, in many aspects, irrelevant since for $N >> 1$, the dBH model  
becomes practically independent of the tunneling term and reduces to a diagonal Hamiltonian in the Fock basis. \\

\begin{figure}[h!]
\centering
\includegraphics[width=\linewidth]{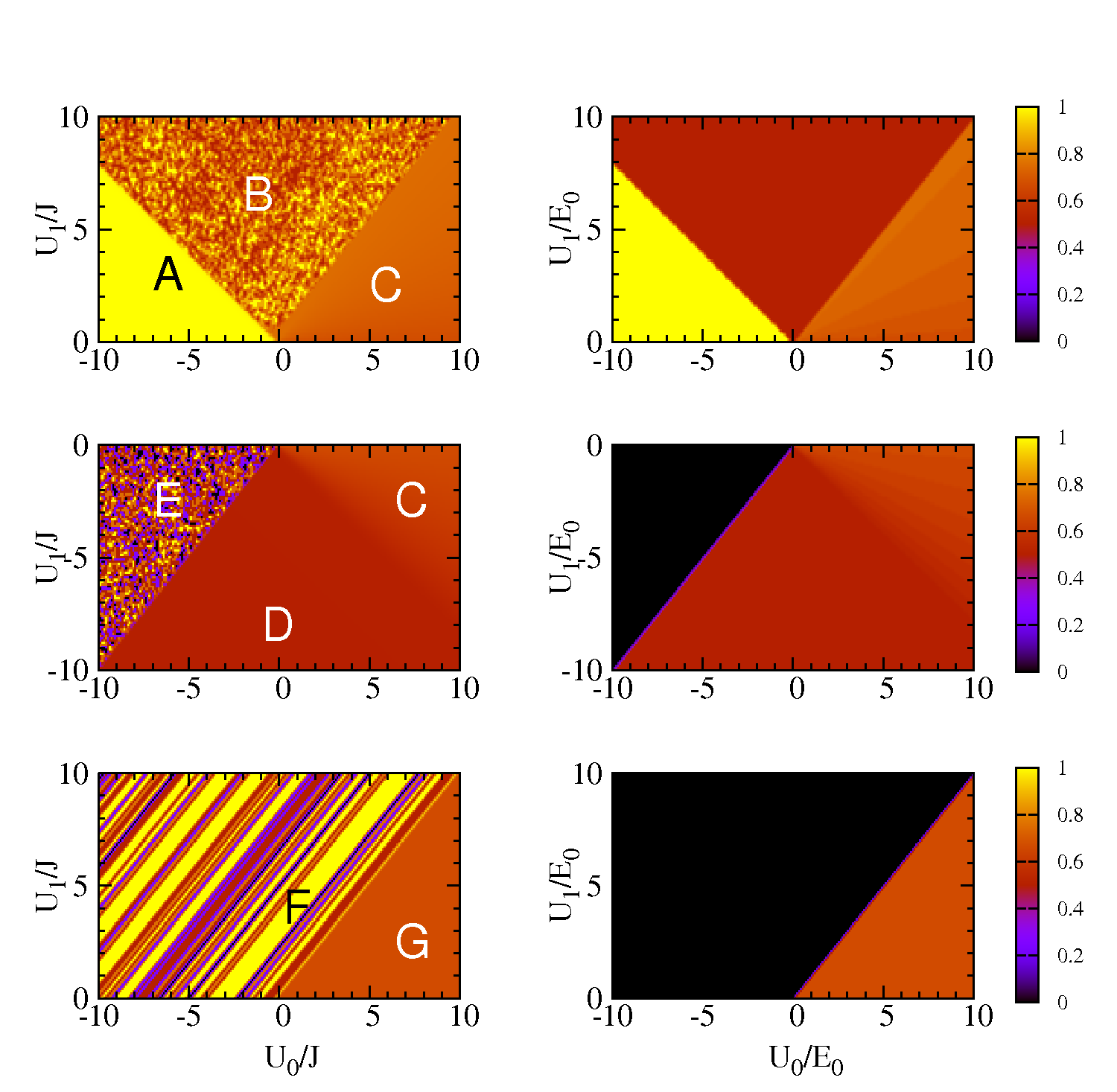}
\caption{
(Color Online) Phase diagram in ($U_{0}/J,U_{1}/J$) plane for the  $\alpha$ (top), $\beta$ (middle) and $\gamma$ (bottom) configurations. Left panels: $(N-\langle \hat{n}_{1}\rangle )/N$ obtained from exact diagonalization of Eq.$\,$(\ref{hamiltonian1}) with $N=48$ bosons. Right panels: boundaries of the phases as predicted in the atomic limit, {\it i.e.\/} for $J=0$. (For the sake of comparison we set $E_0=J$ of the left panels).
}
\label{Fig2}
\end{figure}

The whole phase diagram obtained from the exact diagonalization of the three-site dBH Hamiltonian is displayed in Fig.$\,$\ref{Fig2}. 
It consists of seven different phases. On the left panels, we show the relative vacancy number $(N-\langle \hat{n}_{1}\rangle)/N $ of site 
$1$ versus $U_0$ and $U_1$ for $N=48$ atoms. Phases $A,B$ and $C$ in the $\alpha$-configuration are shown on the top panel, phases 
$C,D$ and $E$ corresponding to the $\beta$-configuration in the middle panel and the two phases $F$ and $G$ of the $\gamma$-configuration 
at the bottom one. All these phases are compared with the ones obtained in the atomic limit by minimizing the energy at $J=0$ (right panels) 
for the corresponding configurations. Inspection of the different phases shows a reasonable agreement between the boundaries in the classical 
limit and the ones obtained by exact diagonalization. Interestingly enough, the exact borders on the phase diagram can be obtained using scaling 
relations derived through entanglement properties \cite{Gallemi2014, DeChiara2012}.
  
Indeed, a first insight on the phase diagram can be obtained in the atomic limit, {\it i.e.} when the tunneling is set to zero. In such case, the Hamiltonian becomes diagonal in the Fock basis and reduces to 
\begin{align}
H_{0}= \;&\hat{n}_{2}(\hat{n}_{1}+\hat{n}_{3})[-U_{0}+U_1\,b\,(1-3cos^2\theta)]+\nonumber \\
+&\;\hat{n}_1\hat{n}_{3}(-U_{0}+U_{1})\,. 
\end{align}
Analytical expressions for the phase boundaries can be obtained by minimizing the energy under the conservation of $N$ and imposing symmetry between wells $1$ and $3$. 
For example, requiring $n_{1}=n_{3}$  yields a mean occupation number:
\begin{equation}
n_{1}= n_{3}=Int \left[ N \frac{ U_{0} -U_{1}b\,(1-3\cos^{2}\theta)}{3U_{0}+U_{1}(1-4b\,(1-3\cos^{2}\theta))}\right],
\label{classic}
\end{equation}         
where $Int[x]$ denotes the integer part of $x\,$. In what it follows we analyze the different configurations separately.

\subsection{$\alpha$-configuration ($U_{1}>0$)}

In the $\alpha$-configuration, the minimization of the energy
leads to three different phases $A,B$ and $C$. Their
borders correspond to the different (integer) occupation values compatible with the constraint $n_{1} = n_{3}$, which for $J = 0$ results in:
\begin{equation}
n_{1}= n_{3}=Int \left[\frac{N}{12} \frac{ 4U_{0}+ 5 U_{1}}{U_{0}+2 U_{1}}\right] 
\label{classic-alpha-1}
\end{equation}

{\it {Phase $A$.}} For $J=0$, this phase extends into the region limited by $0 \le U_{1} \leq -4 U_{0}/5$  and $U_{0} < 0$. In this phase, the inter-site interaction, repulsive between sites $1$-$3$ but attractive otherwise, dominates over the on-site interaction. Exact diagonalization shows indeed the presence of phase $A$, where 
minimization is achieved by accommodating all particles in site 2. The ground state of the system is the product state:
\begin{equation}
\vert \Psi \rangle_A = \vert 0,N,0\rangle\,.
\label{phaseA}
\end{equation} 
\\

{\it {Phase $B$}}. For $J=0$,  a new phase appears in the region  between $0 < -4 U_{0}/5 < U_{1}$ (for $U_{0}< 0$) and  $0 \leq U_{0} \leq U_{1}$ (boundary $B$-$C$). Phase $B$ has a richer structure and appears for both, attractive and repulsive on-site interactions. In contrast to phase $A$, the ground state of the system does not correspond to a single Fock state. For $U_{0}=0$ and $U_{1}\neq 0$ the number of particles in site 1 and 3 must fulfill that $n_{1}=n_{3} < 5/12\,N$.  For $U_{1}=U_{0}$, it follows that $n_{1}=n_{3}=N/4$, and $n_{2}=N/2$, setting the boundary between phases $B$ and $C$.  
The boundaries of phase $B$ have a non trivial dependence on the number of bosons $N$ (see Fig. \ref{Fig3}), and are given by
\begin{equation}
\frac{U_1}{U_0}=4\,\frac{N-2\,n_1-n_3}{14\,n_3+10\, n_1-5\,N}\,.
\label{boundaries-b}
\end{equation}
All the possible values of $n_{1}$ and $n_{3}$ yield the curve of the boundary of phase $B$.  For $J\neq 0$, exact diagonalization (together with the analysis of entanglement), shows that phase $B$ corresponds to a symmetric superposition of sites $1$-$3$, {\it i.e.\/} $1/\sqrt{2}\,(\ket{n_{1},n_{3}}+\ket{n_{3},n_{1}})$, 
conditioned by the occupation number on site $2$:  
\begin{align}
\vert \Psi \rangle_B=& \sum_{a, b} C_{\vert a\vert, b}  \, \frac{1}{\sqrt 2}
   \left[ \, \vert N/2-a, N/2+a-b, b \rangle +\right. \nonumber \\
   &+\left. \vert b, N/2+a-b, N/2-a \rangle \, \right] \,,
\label{phaseB}
\end{align}
with  $-N/2 < a < N/2$, $0<b< N/2+a< N$. The amplitudes  $C_{\vert a\vert, b}$ decrease as the coefficients $|a|$ and $b$ increase.\\

{\it {Phase C.}} At $J=0$, the boundaries of this new phase are given by $U_{1}\le U_{0}$ with $U_{0} \geq 0$.
In this phase, on-site interactions dominate over inter-site ones. When $U_1 = 0$, the dipolar interaction vanishes, the three sites become symmetric and all the particles are distributed equally inside them (equipartition). If we do not set the inter-site interaction to zero and do not fix the occupation number on site 2,  the minimization of the energy together with the conservation of $N$ leads to the following condition for $n_2$:
\begin{equation}
n_{2}=Int \left[ \frac{N}{3}\frac{ U_{0}+ 7/2 \,U_{1}}{U_{0}+2 U_{1}}\right] \,.
\label{phaseBn2}
\end{equation}
Equation (\ref{phaseBn2}) provides the boundaries (as a set of slopes) for each possible occupation number on site 2 
(see also Sec. \ref{sec:entanglementproperties}). The population in this well varies from $N/2$ (for $U_1=U_{0}$) to $N/3$ 
(for $U_{1}=0$).  
For each value of $n_2$ in this range there exists a well defined region in phase $C$ limited by the above boundaries (slopes). 
The transition between these regions is associated to the variation of one particle in site 2.
As will become clear later when analyzing the 
entanglement properties of the states, these different regions within this phase correspond to smooth crossovers.

\subsection{$\beta$-configuration ($U_{1}<0$)}

When all the dipoles are oriented along the horizontal axis three different phases $C, D$ and $E$ appear.\\

{\it {Phase C}}. At $J=0$, $C$ is bounded by  $U_{1}<0$ and $|U_{1}| \leq 5/4 \,U_{0}$. This phase is equivalent to phase $C$ in the $\alpha$-configuration, and again allows different occupation numbers on site $1$ and site $3$. It displays several crossovers connecting now phase $C$ with $D$. The equation equivalent to Eq.$\,$(\ref{phaseBn2}) corresponding to this configuration that sets the boundaries for the different crossovers is now given by:
\begin{equation}
n_{2}=Int \left[ N \left[1- \frac{ 4 U_{0}+ 1/2\, U_{1}}{6 U_{0} + 3 U_{1}}\right] \right] \,,
\label{classic-beta-2}
\end{equation}
where the population on site 2  decreases gradually now from  $n_{2}=N/3$ to $n_{2}=0$.\\

{\it {Phase D}}. At $J=0$, the boundaries of phase $D$ are given by $U_1=-4/5 \,U_0$ (imposing $n_2=0$ in (\ref{classic-beta-2})), and $U_1=U_0$, which is the boundary between phases $E$ and $D$. 
Exact diagonalization shows that phase $D$ has a dominant contribution of the Fock state corresponding to $n_{1}=n_{3}=N/2$, and $n_{2}=0$. It can be understood as the balance between a large attractive inter-site interaction 
between sites 1 and 3, and
the inter-site repulsion between $1$-$2$ and $2$-$3$. Although site $2$ is not correlated to the other ones, the exact ground state is not a product state:
\begin{align}
\vert \Psi \rangle_D &= C_0\, \vert N/2,0,N/2\rangle  \, + \nonumber \\
   &+ \sum_{a} C_{a}  \, \frac{1}{\sqrt 2} \left[ \, \vert N/2-a, 0, N/2 + a \rangle + \right. \nonumber \\
   &+ \left. \vert N/2+ a, 0, N/2-a \rangle \, \right] \,.
\label{phaseD}
\end{align}

{\it {Phase E.}} Finally,  phase $E$ corresponds to a cat state between wells 1 and 3, whereas well 2 is completely empty. 
This ground state can be understood as a balance between a large on-site attraction, a large inter-site attraction between sites $1$-$3$, and the 
symmetry of the geometry.
\begin{equation}
\vert \Psi \rangle_E = \frac{ \vert N,0,0\rangle + \vert 0,0,N\rangle }{\sqrt{2}}  \,.
\label{phaseE}
\end{equation}

\subsection{$\gamma$-configuration ($U_{1}>0$)}

For completeness we also review briefly here the phase diagram when all dipoles are aligned  orthogonal to the plane that contains the three mesoscopic dipoles. As mentioned before (see Table I), in this case the inter-site interaction becomes isotropic, the system acquires a larger symmetry and effectively reduces to a non-dipolar condensate with a renormalized on-site interaction. There are two different phases: $F$ and $G$. 
Phase $F$ is a mesoscopic superposition of the three sites. Minimization of energy leads to the following ground state
  $\ket{\Psi}_{F}= 1/\sqrt{3} \,(\ket{N,0,0}+\ket{0,N,0}+\ket{0,0,N})$, {\it i.e.\/} a maximally entangled state of 3 sites, also called W-state.
 Phase $G$ is a state with equal mean population on all sites. The phase transition occurs just at $U_1=U_0$. 
Since this configuration has no distinctive dipolar effects and has been previously studied \cite{Dell'Anna2013}, 
we shall not address it further.

\begin{figure}[h!]
\begin{center}
\includegraphics[width=\linewidth]{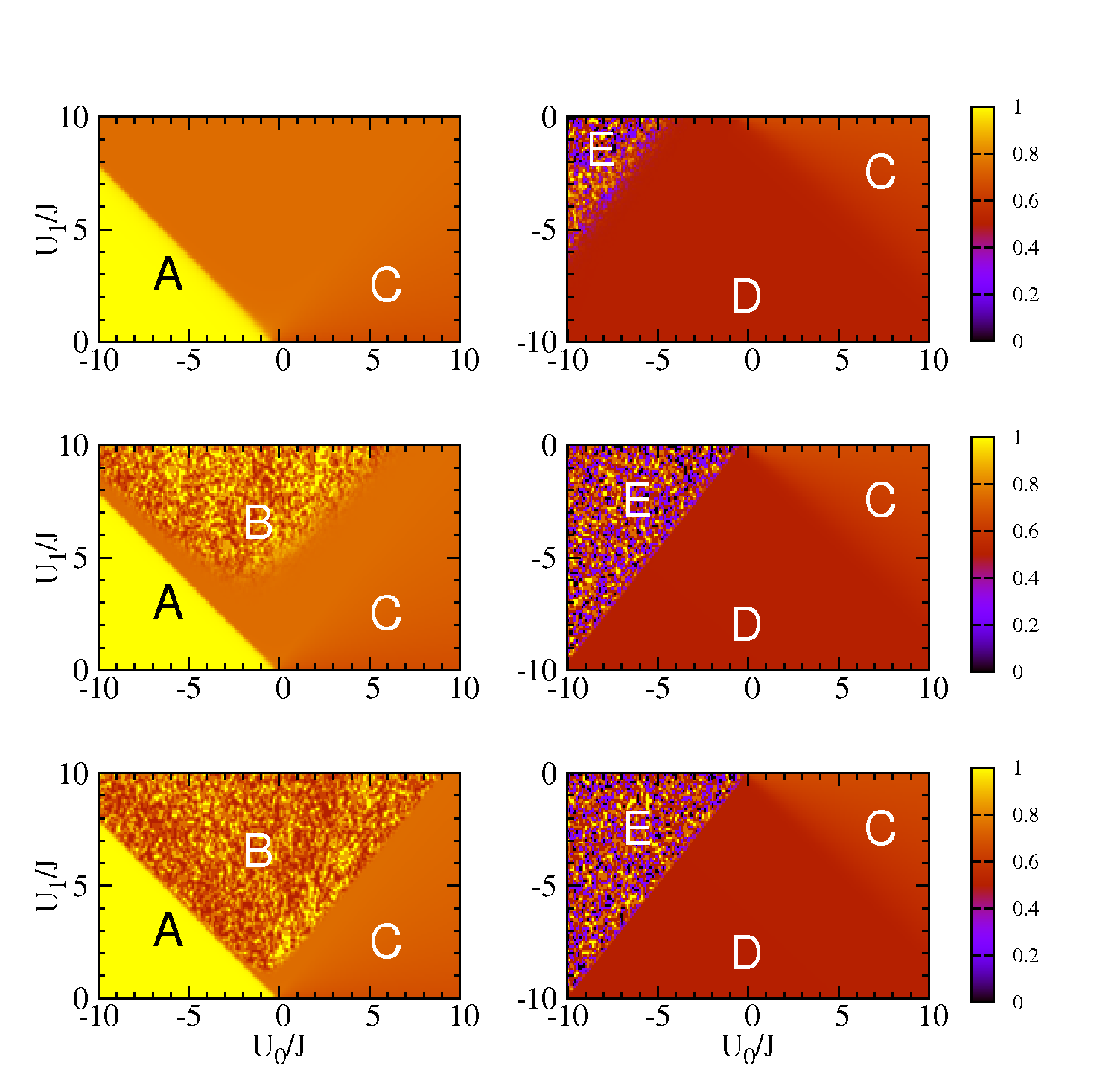}
\caption{(Color online) Phase diagram for the $\alpha$-configuration (left panels) and $\beta$-configuration (right panels) 
obtained from exact diagonalization for $N=12$ (top), $N=24$ (middle) and $N=36$ (bottom) panels.  In the panels, we plot $(N-\langle \hat{n}_{1}\rangle )/N$ versus $U_{0}/J$ and $U_{1}/J$.} 
\label{Fig3}
\end{center}
\end{figure}

Finally, let us comment on the dependence of the phase diagram on the number of bosons $N$.  In Fig.$\,$\ref{Fig3}, we display the phase diagram obtained by exact diagonalization in the $\alpha$ and $\beta$-configurations  
for $N=12,\,24$ and $36$ bosons. All phases, except phase $A$, show a strong dependence on the number of bosons. This can be easily understood since phase $A$ is characterized by the ground state $\ket{\Psi}_{A}=\ket{0,N,0}$, which is a product state, therefore, it shows no correlations between sites. All other phases, which cannot be described as classical states, are very sensitive to quantum fluctuations, thus, the boundaries of the phases depend on the number of bosons \cite{phaseB}. This is reminiscent of the finite size effects shown by strongly correlated systems, where boundaries are only well defined in the thermodynamic limit. Although the number of particles we consider is small, the large $N$ limit ($N\rightarrow \infty$) corresponds here to the regime where tunneling can be neglected and the problem can be treated as three coupled modes in a mean field approach.
\section{Phase analysis and entanglement properties}
\label{sec:entanglementproperties}

A further insight on the quantum phases and phase transitions can be obtained by analyzing entanglement properties. Notice that since our Hamiltonian consists just in three spatial modes (wells) associated to the operators $\hat{a}_{i}$  ($i=1,2,3$), one has to consider entanglement between the different modes and not between the particles. Under this perspective we examine entanglement properties of the ground state of the system for a fixed number of particles, $N$, as a function of the parameters $U_{0}/J$ and $U_{1}/J$.  

\begin{figure}[t!]
\centering
\includegraphics[width=\linewidth]{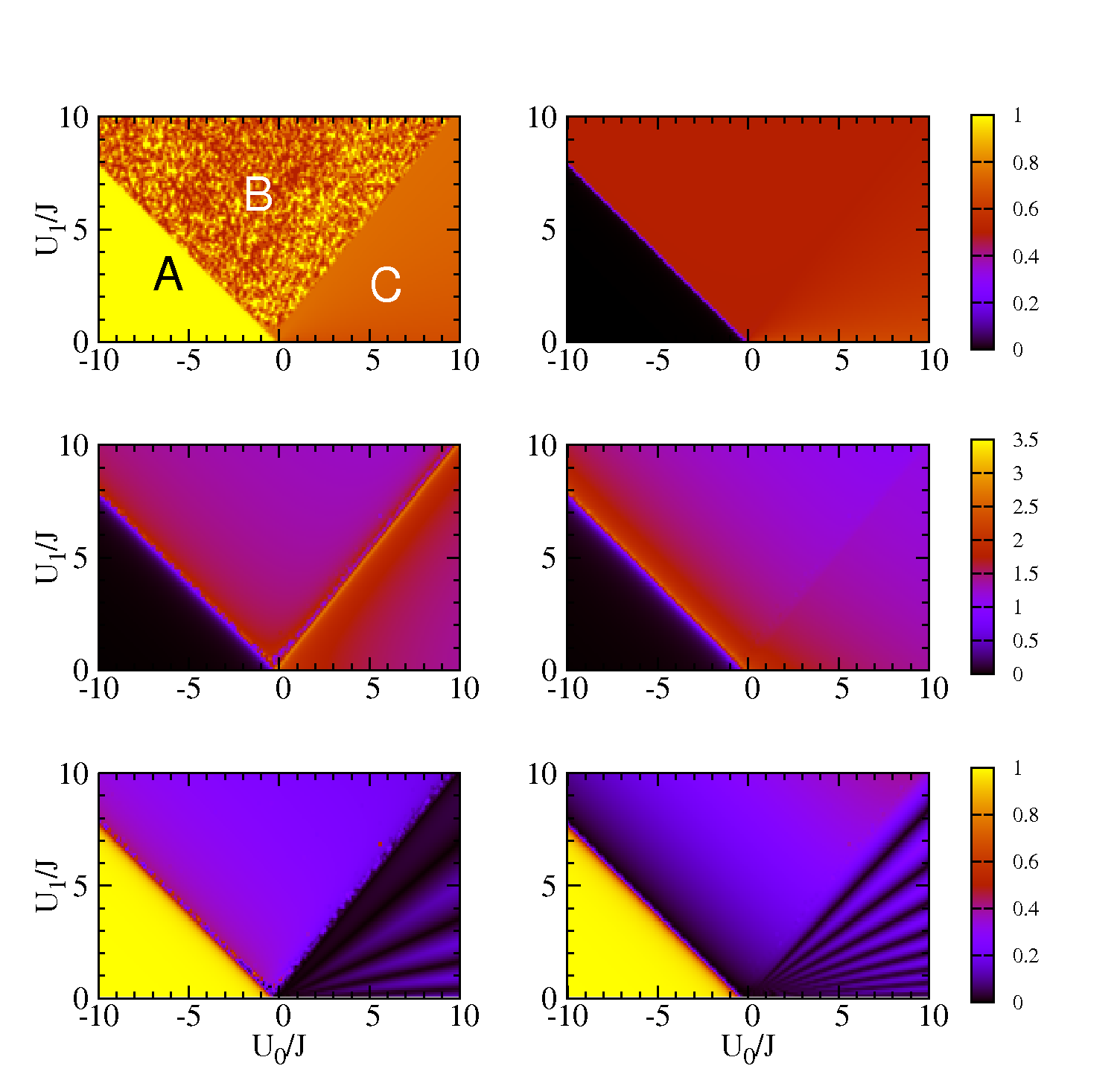}
\caption{ (Color Online) Characterization of the phases $A,B$ and $C$ in the $\alpha$-configuration by means of their phase diagram and entanglement properties: Top panels: $(N-\langle \hat{n}_{1}\rangle )/N$ (left) and $(N-\langle \hat{n}_{2}\rangle )/N$ (right) versus $(U_{0}/J, U_{1}/J)$. Middle panels: von Neumann entropy $S_{1}$ (left) and $S_{2}$ (right). Bottom panels: Schmidt gap $\Delta\lambda^{1}$ (left) and $\Delta\lambda^{2}$ (right). Notice that wells $1$ and $3$ are symmetric for this configuration. The results are obtained by exact diagonalization of Eq.$\,$(\ref{hamiltonian1}) with $N=48$ bosons.}
\label{Fig4}
\end{figure}

\begin{figure}[t!]
\centering
\includegraphics[width=\linewidth]{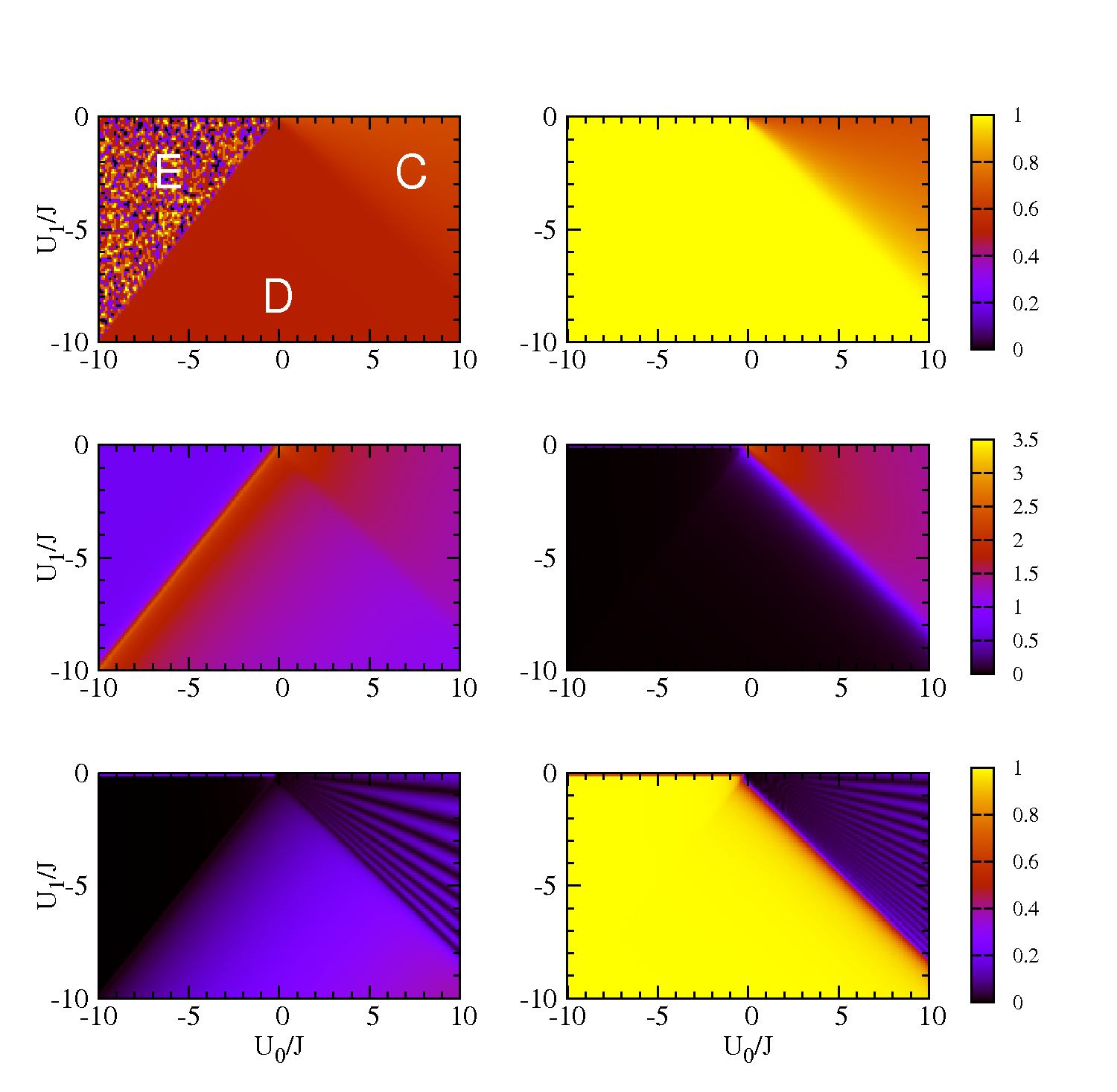}
\caption{ (Color Online) Characterization of the phases $C,D$ and $E$ in the $\beta$-configuration by means of their phase diagram and entanglement properties: Top panels: $(N-\langle \hat{n}_{1}\rangle )/N$ (left) and $(N-\langle \hat{n}_{2}\rangle )/N$ (right) versus $(U_{0}/J, U_{1}/J)$. Middle panels: von Neumann entropy $S_{1}$ (left) and $S_{2}$ (right). Bottom panels: Schmidt gap $\Delta\lambda^{1}$ (left) and $\Delta\lambda^{2}$ (right). Notice that wells $1$ and $3$ are symmetric for this configuration. The results are obtained by exact diagonalization of Eq.$\,$(\ref{hamiltonian1}) with $N=48$ bosons.}
\label{Fig5}
\end{figure}

To determine if a pure state, {\it {e.g}}. the ground state of our system, is or not entangled is trivial: it suffices to check if all of its subsystems are in a pure state, in such a case the ground state is unentangled.  To quantify its entanglement, however, is quite more subtle. Several quantities can be used to quantify entanglement. Among them, the von Neumann ($S(\rho)$), and  Renyi $(S^{n}(\rho))$ entropies are good entanglement monotones for {\it bipartite} splittings. The von Neumann entropy of a subsystem $i$ captures well the fact that for entangled states, the disorder of the subsystem can be greater than the disorder (entropy) of the system itself, something completely forbidden for classical states. The von Neumann entropy $S(\rho_{i})$ has been used to underpin the behavior of the entanglement near criticality in many-body strongly correlated systems. It is defined as:
\begin{equation}
 S_i=S(\rho_{i})=-\mbox{Tr}_i (\rho_i \log \rho_i)=-\sum_{{m}} \lambda^{i}_{m}\log\lambda_{m}^{i}
\end{equation}
where $\rho_i$ is the reduced density matrix (with eigenvalues $\lambda_{i}$) describing subsystem $i$:
\begin{equation}
\rho_i = \mbox{Tr}_j \mbox{Tr}_k \,\rho\,,\,\mbox{with}\,\,\,\rho=\vert \Psi \rangle \langle \Psi \vert\,,
\end{equation}
being $\rho$ the density matrix of the full system. Renyi entropies, defined as $S^{n}=1/(1-n)Tr \rho_{i}^{n}$, are as well non linear 
functions of the eigenvalues of the reduced density matrix. It is instructive to recall that any pure state can be always written in a bipartite
 splitting in the so-called Schmidt decomposition. In our particular case in which we have just 3 different modes {\it i.e.} A, B and C, the ground
 state of the system $\ket{\Psi}_{ABC}$ can be expressed  w.r.t splitting A / BC 
as  $\ket{\Psi}_{ABC}=\sum_{m}\sqrt{\lambda_{m}} \ket{\psi^{m}}_{A}\ket{\phi^{m}}_{BC}$ where $\{\ket{\psi^{m}}_{A}\}$ 
and $\{\ket{\phi^{m}}_{BC}\}$ conform a biorthogonal basis.  The coefficients $\lambda_{i}$, often called the Schmidt coefficients, 
are nothing else than the eigenvalues of the reduced density matrices $\rho_{A}$ and $\rho_{{BC}}$. They fulfill that $\lambda_{m}\ge 0$ 
and $\sum_{m} \lambda_{m}=1$.  Obviously, if the splitting is different, {\it i.e.} AB /C, the biorthogonal basis and their Schmidt coefficients 
change.  There is not an analogue of the Schmidt decomposition for tripartite systems, neither the von Neumann or Renyi entropies are a 
measure of tripartite entanglement. Here, the symmetry imposed by the dipolar interaction on the triple well in the $\alpha$ and $\beta$-configurations 
implies that subsystem 1 and 3 have the same reduced density matrix {\it i.e.} $\rho_{1}= Tr_{23}\rho=\rho_{3}=Tr_{12}\rho$, 
while $\rho_{2}=Tr_{13}\rho$ is different. Therefore, it suffices to consider bipartite splittings. Since there are only two different 
bipartite splittings we obtain two different von Neumann entropies, $S_{1(3)}$ accounting for the entanglement between 
subsystem 1(3) with the rest (1/23), 
and $S_{2}$ reporting the entanglement between subsystem 2 and the rest (2/13). 
Due to this symmetry, which is a consequence of the dipolar anisotropy in these configurations, it is possible to identify the different 
quantum phases by their von Neumann entropy.  In general this does not need to be the case. 




In Fig. \ref{Fig4} and Fig. \ref{Fig5} we display the phase diagram as a function of $U_{1}/J$  and $U_{0}/J$ by means 
of $(N-\langle \hat{n}_{i}\rangle )/N$  for $i=1,2$ (top panels), their corresponding von Neumann entropies  $S_{i}$ (middle panels) 
and their entanglement spectrum using the Schmidt gap $\Delta{\lambda}^{i}=\lambda_{1}^{i}-\lambda_{2}^{i}$, defined as the difference
 between the two largest (and therefore more relevant) non trivially degenerated Schmidt coefficients (the index $i$ refers here to the partition).
 Inspection of the above figures shows a remarkable agreement on the phase diagram obtained either via mean occupation number (top panels), or entanglement properties (middle and bottom panels) with the only exception being the structure displayed by the Schmidt gap for  $U_{0}> 0$. It consist in a series of slopes corresponding to the different crossovers occurring in phase $C$ in both $\alpha$ and $\beta$-configurations, something that neither the mean occupation number or the von Neumann entropy, being averaged quantities, do not show. 
The general traits of the different phases can be now unveiled.  

For product states $\ket{\psi}=\ket{n_{1}}\ket{n_{2}}\ket{n_{3}}$ (phase $A$) the von Neumann entropy of the system is equal to the entropy of its subsystems and identically equal to zero. In such case, the state is already in its Schmidt form with a single Schmidt coefficient equal to one. Any entangled state has necessarily more than one Schmidt coefficient. 

In phase $B$, the single-site entropies $S_{1}$ and $S_{2}$ are: (i) both different from zero, an indication of the presence of entanglement w.r.t any partition and, (ii) larger than $\log2$, ruling out the presence of a single mesoscopic superposition or cat state. The entropy and the Schmidt gap rather indicates that this phase is formed by a superposition of cat states and since $S_2\neq0$ in such phase, the superposition of the cat states is conditioned by the particle occupation of site 2. 

In phase $C$ a rich structure appears in the Schmidt gap which is not present in the entropy or Fock occupation plots. We shall return to this phase later. 

Finally, in phases $E$ and $D$,  the fact that $S_{2}=0$ indicates a ground state of the type $\ket{\psi}=\ket{n_{2}}\ket{\phi_{13}}$ where subsystem 2 is disentangled from the others, but subsystems 1 and 3 are necessarily entangled.  Moreover, since in phase $E$ the von Neumann entropy of sites 1 and 3 is simply $\log2$, one can deduce the existence of a mesoscopic quantum superposition of two terms {\it i.e.} $1/\sqrt{2}\, (\ket{N,0,0}+\ket{0,0,N})$. This is further supported by the value of the Schmidt gap $\Delta\lambda^{1}=0$ meaning that the two first eigenvalues of the Schmidt decomposition of this partition are equal.  This is not the case for phase $D$, where the entropies are larger and the Schmidt gap different from zero. 
\begin{figure}[h!]
\centering
\includegraphics[width=\linewidth]{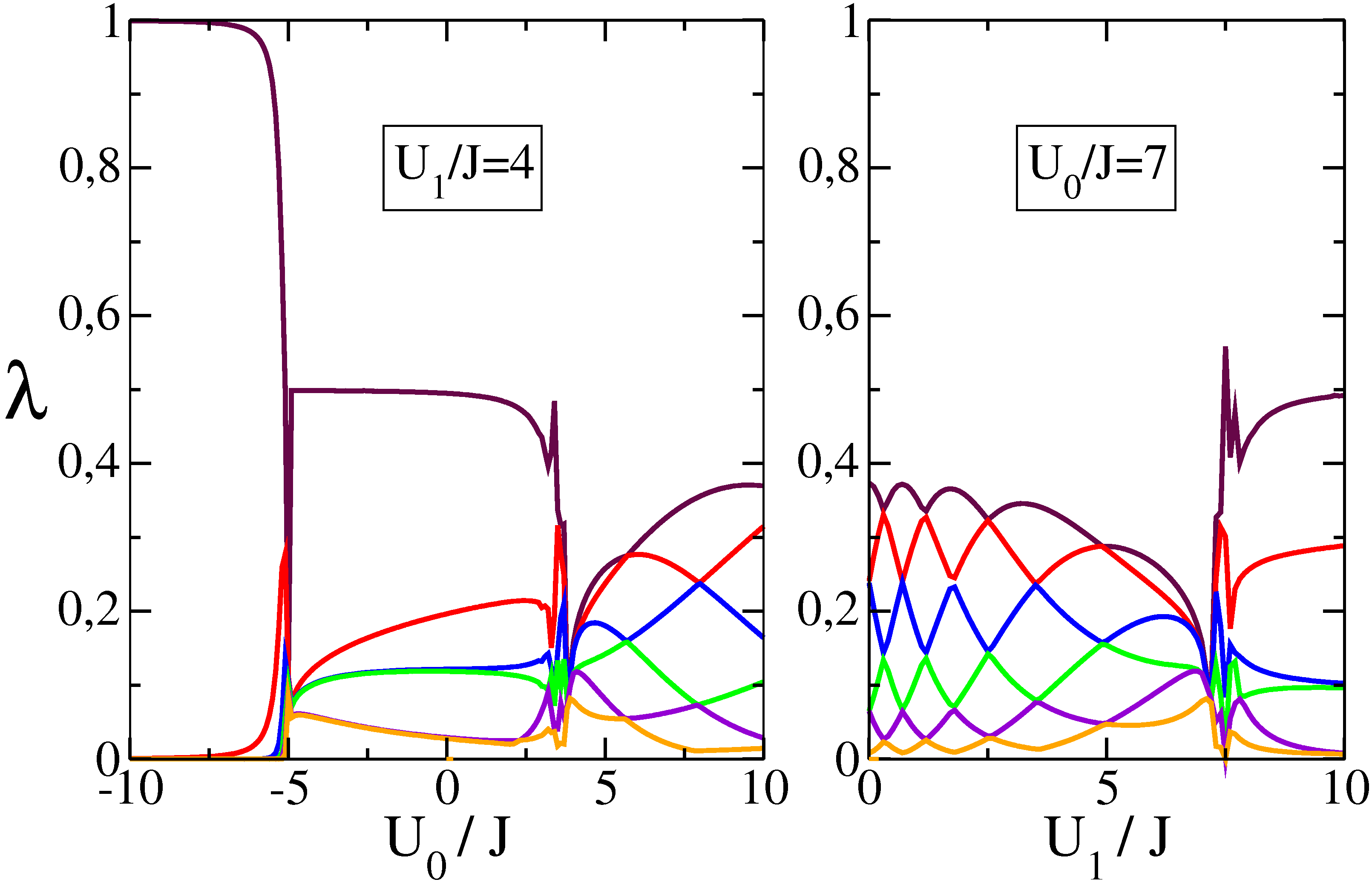}
\caption{
Entanglement spectrum (first six values in decreasing order) in the $\alpha$-configuration for $U_1/J=4$ (left) and for a fixed value of $U_{0}/J=7$ (right). Quantum phase transitions correspond to the points in which degeneracy between the different eigenvalues occurs. Crossovers are associated to crossings between the different Schmidt coefficients.
}
\label{Fig6}
\end{figure}
Boundaries between quantum phases are normally related to the presence of criticality, a meaningful concept in the thermodynamic limit. For instance, in spin chain systems, it can be observed that when approaching a quantum phase transition the entanglement spectrum tends to a band with all eigenvalues $\lambda_{m}$ quasi degenerated. This critical behavior, where fluctuations are large, is associated with the fact that entanglement should be present at all length-scales.  It has also been shown that for finite spin chain systems, outside the thermodinamic limit, the Schmidt gap closes normally faster than the other values, that is, the largest eigenvalues become faster degenerated \cite{DeChiara2012}. 

To better understand the structure presented in the previous panels we display in Fig.$\,$\ref{Fig6}, the six first eigenvalues (in decreasing order) along a cut of the phase diagram. Thus, in the left plot of Fig.$\,$\ref{Fig6} we show the entanglement spectrum for a fixed value of  $U_{1}/J=4$,
while in the right plot the entanglement spectrum is represented for $U_{0}/J=7$. As can be seen in the plots of the entanglement spectrum, phase boundaries are  associated to a collapse of all the eigenvalues. Thus, for $U_{1}/J=4$ degeneracy of the eigenvalues appears when approaching $U_{0}/J=-5$, corresponding to the phase transition $A$-$B$, and again approaching $U_{0}/J=4$ where the transition $B$-$C$ appears. Moreover, for larger values of $U_{0}/J$ there is not a collapse of the eigenvalues but  a crossing of them at few points. Even more clear is this crossing of eigenvalue pairs in the right plot of Fig.$\,$\ref{Fig6} to finally collapse near $U_{1}/J=7$ where the transition $C$-$B$ is now crossed. The information encoded in the crossings is related to changes in the number of bosons in well 2 and, as a consequence, a change in the Fock states participating in the ground state.\\

\section{Summary and conclusions}

We have shown that mesoscopic samples of polarized dipolar atoms confined in three spatially separated traps realize different quantum phases depending on the dipole orientation. Using for their description an extended dipolar Bose-Hubbard Hamiltonian, the competition set by the anisotropy of the inter-site dipole-dipole interaction, together with on-site interaction strongly determines the possible phases of the ground states. We have solved the dBH Hamiltonian by means of exact diagonalization for different number of bosons, and shown in which cases the ground state is compatible with a mean field description. To this aim we have further analyzed the properties of each phase by means of their entanglement properties, going beyond averaged quantities like the von Neumann entropy to discuss and distinguish the differences between phase boundaries and crossovers in these setups. The former are accompanied by a flat band distribution of all Schmidt coefficients while the latter are shown as a crossover between different Schmidt coefficients without full degeneracy.  A more detailed study on the scaling behavior of the entanglement in these systems, allowing for the determination of critical quantities and exponents is beyond the scope of the present work and will be presented in the future \cite{Gallemi2014}.

\begin{acknowledgements}
We acknowledge financial support from the Spanish MINECO
(FIS2008-01236 and FIS2011-28617-C02-01) and the European Regional development Fund, Generalitat de Catalunya Grant No. 
SGR2009-00347 and Grant No. SGR2009-1289.
\end{acknowledgements}


\thebibliography{99}

\bibitem{Griesmaier2005} 
A. Griesmaier, J. Werner, S. Hensler, J. Stuhler, and T. Pfau, 
Phys. Rev. Lett. \textbf{94}, 160401 (2005).
%
\bibitem{Lu2011} 
M. Lu, N. Q. Burdick, S. H. Youn, and B. L. Lev, 
Phys. Rev. Lett. \textbf{107}, 190401 (2011).
%
\bibitem{Aikawa2012}
K. Aikawa, A. Frisch, M. Mark, S. Baier, A. Rietzler, R. Grimm, and F. Ferlaino,
Phys. Rev. Lett. \textbf{108}, 210401 (2012).
%
\bibitem{Deiglmayr2008}
J. Deiglmayr, A. Grochola, M. Repp, K. M\"ortlbauer, C. Gl\"uck,
J. Lange, O. Dulieu, R. Wester, and M. Weidem\"uller, Phys. Rev.
Lett. \textbf{101}, 133004 (2008)
%
\bibitem{DeMiranda2011}
M. H. G. de Miranda, A. Chotia, B. Neyenhuis, D. Wang, G. Qu\'em\'ener, S. Ospelkaus, J. L. Bohn, J. Ye, D. S. Jin, 
Nat. Phys. \textbf{7}, 502 (2011).
%
\bibitem{Lahaye2009} 
T. Lahaye, C. Menotti, L. Santos, M. Lewenstein, and T. Pfau, 
Rep. Prog. Phys. \textbf{72}, 126401 (2009).
%
\bibitem{Abad2010} 
M. Abad, M. Guilleumas, R. Mayol, M. Pi, and D. M. Jezek, 
Phys. Rev. A. \textbf{81}, 043619 (2010).
%
\bibitem{Albiez2005}
M. Albiez, R. Gati, J. F\"olling, S. Hunsmann, M. Cristiani and M. K. Oberthaler, 
Phys. Rev. Lett, {\bf 95}, 010402 (2005).
%
\bibitem{Levy2007}
S. Levy, E. Lahoud, I. Shomroni and J. Steinhauer, 
Nature, {\bf 449}, 579 (2007).
%
\bibitem{Zin2008} 
P. Zi\'n, J. Chwede\'nczuk, B. Ole\'s, K. Sacha, and M. Trippenbach, 
Eur. Phys. Lett. \textbf{83}, 64007 (2008).
%
\bibitem{Javanainen1986} 
J. Javanainen, M. Y. Ivanov, 
Phys. Rev. A, \textbf{60}, 2351 (1999).
%
\bibitem{Buonsante2012} 
P. Buonsante, R. Burioni, E. Vescovi, A. Vezzani,
Phys. Rev. A, \textbf{85}, 043625 (2012).
%
\bibitem{JuliaDiaz2010}
B. Juli\'a-D\'iaz, D. Dagnino, M. Lewenstein, J. Martorell and A. Polls, 
Phys. Rev. A, {\bf 81}, 023615 (2010).
%
\bibitem{Sakmann2009}
K. Sakmann, A. I. Streltsov, O. E. Alon and L. S. Cederbaum, 
Phys. Rev. Lett, {\bf 103}, 220601 (2009).
%
\bibitem{Abad2011} 
M. Abad, M. Guilleumas, R. Mayol, M. Pi, and D. M. Jezek, 
Eur. Phys. Lett. \textbf{94}, 10004 (2011).
%
\bibitem{Lahaye2010} 
T. Lahaye, T. Pfau and L. Santos,
Phys. Rev. Lett.  \textbf{104}, 170404 (2010).
%
\bibitem{Zhang2012}
A. X. Zhang and J. K. Xue, 
J. Phys. B: At. Mol. Opt. Phys. {\bf 45}, 145305 (2012).
%
\bibitem{Peter2012}
D. Peter, K. Pawlowski, T. Pfau, and K. Rzazewski,
J. Phys. B: At. Mol. Opt. Phys. {\bf 45}, 225302 (2012).
%
\bibitem{Dell'Anna2013}
L. Dell'Anna, G. Mazzarella, V. Penna, and L. Salasnich, 
Phys. Rev. A. \textbf{87}, 053620 (2013).
%
\bibitem{limitation}
This condition sets an important limitation on the validity of deriving a Bose-Hubbard model from 
the many-body Hamiltonian \cite{Peter2012}.
%
\bibitem{Gallemi2014}
A. Gallem\'{\i}, M. Guilleumas, R. Mayol, and A. Sanpera, in preparation.
%
\bibitem{DeChiara2012}
G. De Chiara, L. Lepori, M. Lewenstein and A. Sanpera, 
Phys. Rev. Lett {\bf 109}, 237208 (2012).
%
\bibitem{phaseB}
The boundaries of phase $B$ change dramatically with the number of bosons. For instance, in the case of $N=12$, 
they appear outside the range of the plot.
%
\end{document}